\documentclass[multphys,vecphys,sprmidx]{svmult}

\usepackage{amsmath}
\usepackage{amssymb}
\usepackage{makeidx}     
\usepackage{graphicx}    
\usepackage{multicol}    

\begin{document}

\title*{Role of defects and disorder in the half-metallic full-Heusler compounds}

\author{Iosif Galanakis\inst{1} \and  Kemal \"Ozdo\~gan \inst{2} \and
Ersoy \c Sa\c s\i o\~glu\inst{3,4}}

\institute{Department of Materials Science, School of Natural
Sciences, University of Patras,  GR-26504 Patra, Greece \and
Department of Physics, Gebze Institute of Technology, Gebze,
41400, Kocaeli, Turkey \and Institut f\"ur Festk\"orperforschung,
Forschungszentrum J\"ulich, D-52425 J\"ulich, Germany \and Fatih
University,  Physics Department, 34500,    B\" uy\" uk\c cekmece,
\.{I}stanbul, Turkey}

\maketitle

\textit{Emails: galanakis@upatras.gr, kozdogan@gyte.edu.tr,
e.sasioglu@fz-juelich.de}

\begin{abstract}
Half-metallic ferromagnets and especially the full-Heusler alloys
containing Co are at the center of scientific research due to
their potential applications in spintronics. For realistic devices
it is important to control accurately the creation of defects in
these alloys. We review some of our late results on the role of
defects and impurities in these compounds. More precisely we
present results for the following cases (i) doping and disorder in
Co$_2$Cr(Mn)Al(Si) alloys, (ii) half-metallic ferrimagnetism
appeared due to the creation of Cr(Mn) antisites in these alloys,
(iii) Co-doping in Mn$_2$VAl(Si) alloys leading to half-metallic
antiferromagnetism, and finally (iv) the occurrence of vacancies
in the full-Heusler alloys containing Co and Mn. These results are
susceptible of encouraging further theoretical and experimental
research in the properties of these compounds.
\end{abstract}

\section{Introduction}

Heusler alloys \cite{heusler} have attracted during the last
century a great interest due to the possibility to study in the
same family of alloys a series of interesting diverse magnetic
phenomena like itinerant and localized magnetism,
antiferromagnetism, helimagnetism etc
\cite{landolt,landolt2,Pierre97,Tobola1,Tobola2,Tobola3}. The
first Heusler alloys studied were crystallizing in the $L2_1$
structure which consists of 4 fcc sublattices. Afterwards, it was
discovered that it is possible to leave one of the four
sublattices unoccupied ($C1_b$ structure). The latter compounds
are often called half- or semi-Heusler alloys, while the $L2_1$
compounds are referred to as full-Heusler alloys. NiMnSb belongs
to the half-Heusler alloys \cite{Watanabe1976}. In  1983 de Groot
and his collaborators \cite{groot} showed by using
first-principles electronic structure calculations that this
compound is in reality half-metallic, i.e. the minority band is
semiconducting with a gap at the Fermi level $E_F$, leading to
100\% spin polarization at $E_F$. Other known half-metallic
materials except the half- and full-Heusler alloys (see
\cite{book,Review1,Review2,Review3} and references therein) are
some oxides (\textit{e.g} CrO$_2$ and Fe$_3$O$_4$)
\cite{Soulen98}, the manganites (\textit{e.g}
La$_{0.7}$Sr$_{0.3}$MnO$_3$) \cite{Soulen98}, the double
perovskites (\textit{e.g.} Sr$_2$FeReO$_6$) \cite{Kato}, the
pyrites (\textit{e.g} CoS$_2$) \cite{Pyrites}, the transition
metal chalcogenides (\textit{e.g} CrSe) and pnictides
(\textit{e.g} CrAs) in the zinc-blende or wurtzite structures
\cite{GalanakisZB1,GalanakisZB2,GalanakisZB3,GalanakisZB4,Xie1,Xie2,Xie3,Xie4,Xie5,Xie6,Xie7,Xie8,Xie9,Xie10,Xie11,Akinaga1,Akinaga2,Akinaga3,Akinaga4,Akinaga5,Akinaga6,Akinaga7,Akinaga8,Zhao1,Zhao2},
the europium chalcogenides (\textit{e.g} EuS) \cite{Temmerman} and
the diluted magnetic semiconductors (\textit{e.g} Mn impurities in
Si or GaAs) \cite{FreemanMnSi,Akai98}.
 Although thin films of CrO$_2$
and La$_{0.7}$Sr$_{0.3}$MnO$_3$ have been verified to present
practically 100\% spin-polarization at the Fermi level at low
temperatures \cite{Soulen98,Park98}, the Heusler alloys remain
attractive for technical applications like spin-injection devices
\cite{Data}, spin-filters \cite{Kilian00}, tunnel junctions
\cite{Tanaka99}, or GMR devices \cite{Caballero98,Hordequin} due
to their relatively  high  Curie temperature compared to these
compounds \cite{landolt}.

Ishida and collaborators studied by means of \textit{ab-initio}
calculations the full-Heusler compounds of the type Co$_2$MnZ,
where Z stands for Si and Ge, and have shown that they are
half-metals \cite{Ishida}. Later the origin of half-metallicity in
these compounds has been largely explained \cite{GalaFull}. Many
experimental groups during the last years have worked on these
compounds and have tried to synthesize them mainly in the form of
thin films and incorporate them in spintronic devices.  The group
of Westerholt  has extensively studied the properties of
Co$_2$MnGe films and they have incorporated this alloy in the case
of spin-valves and multilayer structures
\cite{Westerholt1,Westerholt2,Westerholt3}. The group of Reiss
managed to create magnetic tunnel junctions based on Co$_2$MnSi
\cite{Reiss1,Reiss2}. A similar study of Sakuraba and
collaborators resulted in the fabrication of magnetic tunnel
junctions using Co$_2$MnSi as one magnetic electrode and Al-O as
the barrier (Co$_{75}$Fe$_{25}$ is the other magnetic electrode)
and their results are consistent with the presence of
half-metallicity for Co$_2$MnSi \cite{Sakuraba}. Dong and
collaborators recently managed to inject spin-polarized current
from Co$_2$MnGe into a semiconducting structure \cite{Dong}.
Finally Kallmayer \textit{et al.} studied the effect of
substituting Fe for Mn in Co$_2$MnSi films and have shown that the
experimental extracted magnetic spin moments are compatible with
the half-metallicity for small degrees of doping \cite{Kallmayer}.

It is obvious from the experimental results that the full-Heusler
compounds containing Co and Mn are of particular interest for
spintronics. Not only they combine high Curie temperatures and
coherent growth on top of semiconductors (they consist of four fcc
sublattice with each one occupied by a single chemical element)
but in real experimental situations they can preserve a high
degree of spin-polarization at the Fermi level. In order to
accurately control their properties it is imperative to
investigate the effect of defects, doping and disorder on their
properties. Recently Picozzi \textit{et al.} published a study on
the effect of defects in Co$_2$MnSi and Co$_2$MnGe \cite{Picozzi}
followed by an extensive review of the defects in these alloys
\cite{PicozziReview}.

Authors have studied in the recent years several aspects of these
half-metallic alloys like the origin of the gap
\cite{GalaFull,GalaHalf}, properties of surfaces
\cite{Surf1,Surf2,Surf3} and interfaces with semiconductors
\cite{Inter1,Inter2}, the quaternary \cite{GalanakisQuart,JAP},
the orbital magnetism \cite{Orbit1,Orbit2}, the exchange constants
\cite{ExchConst}, the magneto-optical properties \cite{Gala-XMCD},
the half-metallic ferrimagnetic Heusler alloys like Mn$_2$VAl,
\cite{Mn2VZ,Mn2VZ-2} and the fully-compensated half-metallic
ferrimagnets or simply half-metallic antiferromagnets
\cite{Cr2MnZ}.  In this chapter we will overview some of our
results on the defects in the half-metallic full Heusler alloys
obtained using the full--potential nonorthogonal local--orbital
minimum--basis band structure scheme (FPLO)
\cite{koepernik1,koepernik2} within the local density
approximation (LDA) \cite{LDA1,LDA2,LDA3} and employing the
coherent potential approximation (CPA) to simulate the disorder in
a random way \cite{koepernik2}. In section \ref{sec2} we present
the physics of defects in the ferromagnetic Heusler alloys
containing Co and Mn like Co$_2$MnSi \cite{APL,PRB-def}. In
section \ref{sec3} we show the creation of half-metallic
ferrimagnets based on the creation of Cr and Mn antisites in
Co$_2$(Cr or Mn)(Al or Si) alloys \cite{PSS-RRL,SSC} and in
section \ref{sec4} we expand this study to cover the case of Co
defects in ferrimagnetic Mn$_2$VAl and Mn$_2$VSi alloys leading to
half-metallic antiferromagnets \cite{PRB-HMA}. In section
\ref{sec5} we investigate a special case: the occurrence of
vacancies in the full-Heusler compounds \cite{PSS-RRL2}. Finally
in section \ref{sec6} we summarize and conclude.

\begin{figure}
\includegraphics[width=\textwidth]{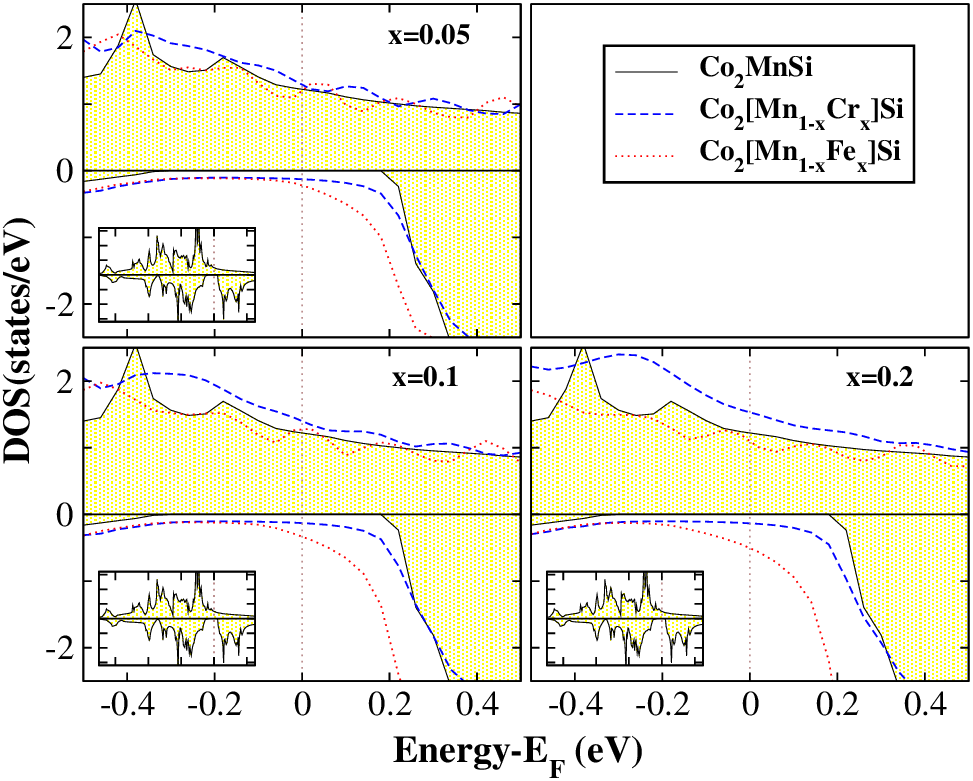}
\caption{(Color online) Spin-resolved total density of states
(DOS) for the case of Co$_2$[Mn$_{1-x}$Cr$_x$]Si and
Co$_2$[Mn$_{1-x}$Fe$_x$]Si for three difference values of the
doping concentration $x$. DOS's are compared to the one of the
undoped Co$_2$MnSi alloy. In the insets we present the DOS for a
wider energy range. We have set the Fermi level as the zero of the
Energy axis. Note that positive values of DOS refer to the
majority-spin electrons and negative values to the minority-spin
electrons.} \label{fig1}
\end{figure}

\section{Defects in full-Heuslers containing Co and Mn \label{sec2} }

We will start our discussion presenting our results on the defects
in the case of Co$_2$MnZ alloys where Z is Al and Si
\cite{APL,PRB-def}. \noindent The first part of our investigation
in this section concerns the doping of Co$_2$MnSi. To simulate the
doping by electrons we substitute Fe for Mn while to simulate the
doping of the alloys with holes we substitute Cr for Mn. We have
studied the cases of moderate doping substituting 5\%, 10\% and
20\% of the Mn atoms. We will start our discussion from figure
\ref{fig1} where we present the total density of states (DOS) for
the Co$_2$[Mn$_{1-x}$Fe$_x$]Si and Co$_2$[Mn$_{1-x}$Cr$_x$]Si
compounds. As discussed in reference \cite{GalaFull} the gap is
created between states located exclusively at the Co sites. The
states low in energy (around -6 eV) originate from the low-lying
$p$-states of the $sp$ atoms (there is also an $s$-type state very
low in energy which is not shown in the figure). The majority-spin
occupied states form a common Mn-Co band while the occupied
minority states are mainly located at the Co sites and the
minority unoccupied states at the Mn sites. Doping the perfect
ordered alloy with either Fe or Cr smoothens the valleys and peaks
along the energy axis. This is a clear sign of the chemical
disorder; Fe and Cr induce peaks at slightly different places than
the Mn atoms resulting to this smoothening and as the doping
increases this phenomenon becomes more intense. The important
detail is what happens around the Fermi level and in what extent
is the gap in the minority band affected by the doping. So now we
will concentrate only at the enlarged regions around the Fermi
level. The blue dashed lines represent the Cr-doping while the red
dash-dotted lines are the Fe-doped alloys. Cr-doping has only
marginal effects to the gap. Its width is narrower with respect to
the perfect compounds but overall the compounds retain their
half-metallicity. In the case of Fe-doping the situation is more
complex. Adding electrons to the system means that, in order to
retain the perfect half-metallicity, these electrons should occupy
high-energy lying antibonding majority states. This is
energetically not very favorable and for these moderate degrees of
doping a new shoulder appears in the unoccupied states which is
close to the right-edge of the gap; a sign of a large change in
the competition between the exchange splitting of the Mn majority
and minority states and of the Coulomb repulsion. In the case of
the 20\% Fe doping this new peak crosses the Fermi level and the
Fermi level is no more exactly in the gap but slightly above it.
Further substitution should lead to the complete destruction of
the half-metallicity as in the Quaternary Heusler alloys with a
Mn-Fe disordered site \cite{GalaQuat}. Recent ab-initio
calculations including the on-site Coulomb repulsion (the
so-called Hubbard $U$) have predicted that Co$_2$FeSi is in
reality half-metallic reaching a total spin magnetic moment of 6
$\mu_B$ which is the largest known spin moment for a half-metal
\cite{Kandpal,Wurmehl}.

We expand our theoretical work to include also the case of
Co$_2$MnAl compound which has one valence electron less than
Co$_2$MnSi. The extra electron in the the latter alloy occupies
majority states leading to an increase of the exchange splitting
between the occupied majority and the unoccupied minority states
and thus to larger gap-width for the Si-based compound. In the
case of Al-based alloy the bonding and antibonding minority
d-hybrids almost overlap and the gap is substituted by a region of
very small minority density of states (DOS); we will call it a
pseudogap. In both cases the Fermi level falls within the gap
(Co$_2$MnSi) or the pseudogap (Co$_2$MnAl) and an almost perfect
spin-polarization at the Fermi level is preserved.

\begin{figure}
\includegraphics[width=\textwidth]{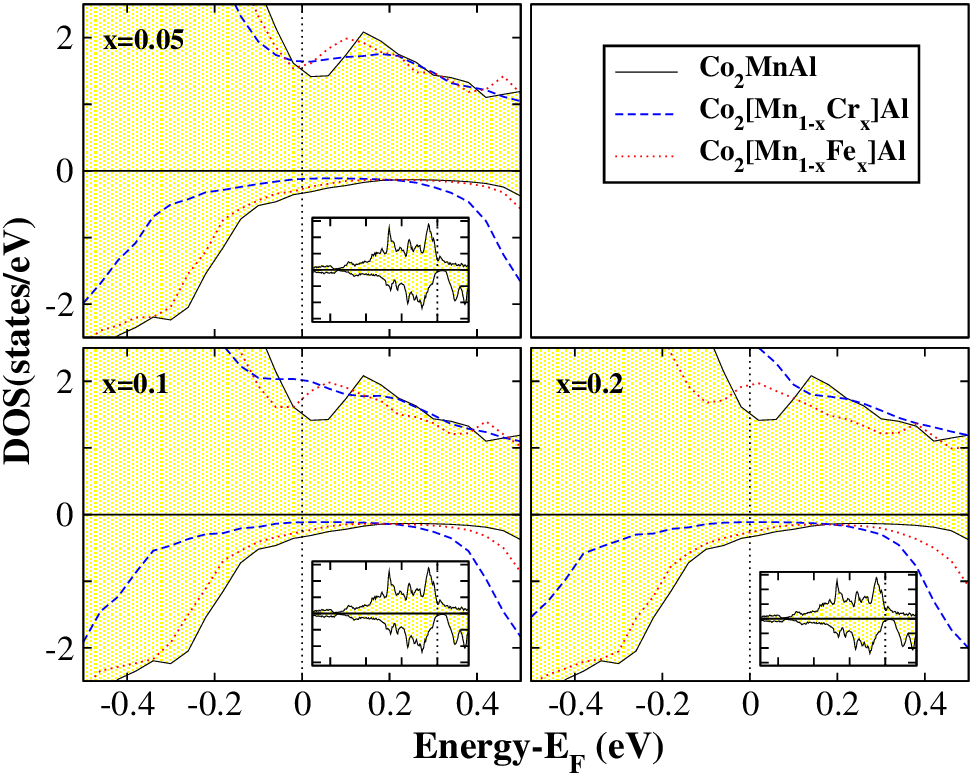}
\caption{(Color online) Spin-resolved DOS for the case of
Co$_2$[Mn$_{1-x}$Cr$_x$]Al and Co$_2$[Mn$_{1-x}$Fe$_x$]Al for
three values of the doping concentration $x$. DOS's are compared
to the one of the undoped Co$_2$MnAl alloy. Details as in figure
\ref{fig1}. \label{fig2}}
\end{figure}

We substitute either Fe or Cr for Mn to simulate the doping by
electrons and holes, respectively, in Co$_2$MnAl.   In figure
\ref{fig2} we present the total density of states (DOS) for the
Co$_2$[Mn$_{1-x}$(Fe or Cr)$_x$]Al alloys to compare to the
perfect Co$_2$MnAl alloys.  As was the case also for the compounds
in reference \cite{APL} and discussed above the majority-spin
occupied states form a common Mn-Co band while the occupied
minority states are mainly located at the Co sites and minority
unoccupied at the Mn sites (note that the minority unoccupied
states near the gap are Co-like but overall the Mn-weight is
dominant). The situation is reversed with respect to the
Co$_2$MnSi compound and Cr-doping has significant effects on the
pseudogap. Its width is larger with respect to the perfect
compound and becomes slightly narrower as the degree of the doping
increases. Fe-doping on the other hand almost does not change the
DOS around the Fermi level. The extra-electrons occupy high-energy
lying antibonding majority states but since Co$_2$MnAl has one
valence electron less than Co$_2$MnSi half-metallicity remains
energetically favorable and no important changes occur upon
Fe-doping and   further substitution of Fe for Mn should retain
the half-metallicity even for the Co$_2$FeAl compound although
LDA-based ab-initio calculations predict that the limiting case of
Co$_2$FeAl is almost half-metallic \cite{GalaQuat}.

\begin{table}
\caption{Total and atom-resolved spin magnetic moments for the
case of excess of Mn ($x$ positive) or $sp$ atoms ($x$ negative)
atoms in $\mu_B$. The spin moments have been scaled to one atom.
In the second column the ideal total spin moment if the compound
was half-metallic.} \label{table1}
 \begin{tabular}{r|c|cccc}
 \hline\noalign{\smallskip}
 & & \multicolumn{4}{c}{Co$_2$Mn$_{1+x}$Al$_{1-x}$} \\ \hline
  $x$ & Ideal &   Total  &   Co &   Mn&  Al   \\ \hline
 -0.20  &  3.40 & 3.26 &   1.09&   2.89  &  -0.12\\
 -0.10  &  3.70 & 3.64 &   1.22&   2.84  &  -0.13\\
 -0.05  &  3.85 & 3.83 &   1.29&   2.83  &  -0.13\\
  0.00  &  4.00 & 4.04 &   1.36&   2.82  &  -0.14\\
   0.05  &  4.15 & 4.22 &   1.40&   2.81  &  -0.14\\
  0.10  &  4.30 & 4.40 &   1.44&   2.81  &  -0.14\\
  0.20  &  4.60 & 4.80 &   1.54&   2.81  &  -0.15 \\ \hline
 & & \multicolumn{4}{c}{Co$_2$Mn$_{1+x}$Si$_{1-x}$} \\ \hline
  $x$ & Ideal &   Total  &   Co &   Mn&  Al  \\ \hline
  -0.20  & 4.40&   4.40&   1.92 &  3.19  &  -0.06   \\
 -0.10  & 4.70&   4.70&   1.95    & 3.15 &    -0.08  \\
 -0.05  & 4.85&   4.85&   1.96  & 3.14  &  -0.08     \\
   0.00 & 5.00&   5.00&  1.96 &   3.13 &-0.09  \\
  0.05  & 5.15&   5.15&   1.99  & 3.10  &  -0.10  \\
  0.10  & 5.30&   5.30&  2.00   & 3.09    & -0.10  \\
  0.20  & 5.60&   5.60&    2.03  &  3.05 &    -0.11  \\
   \noalign{\smallskip}\hline
\end{tabular}
\end{table}

Finally we shall briefly discuss the case of disorder simulated by
the excess of the Mn or the $sp$ atoms. In table \ref{table1} we
have gathered the total and atomic spin moments for all cases
under study. Substituting 5\%, 10\%, 15\% or 20\% of the Mn atoms
by Al or Si ones, corresponding to the negative values of $x$ in
the table, results in a decrease of 0.15, 0.30, 0.45 and 0.60 of
the total number of valence electrons in the cell, while the
inverse procedure results to a similar increase of the mean value
of the number of valence electrons. Contrary to the Si compound
which retains the perfect half-metallicity, the Al-based compound
is no more half-metallic. In the case of Co$_2$MnSi disorder
induces states at the edges of the gap keeping the half-metallic
character but this is no more the case for Co$_2$MnAl compound
where no real gap exists \cite{APL,PRB-def}.

\section{Defects driven half-metallic ferrimagnetism \label{sec3}}

In the previous section we have examined the case of defects in
half-metallic ferromagnets. But  the ideal case for applications
would be a half-metallic antiferromagnet (HMA), also known as
fully-compensated ferrimagnet  \cite{Leuken}, since such a
compound would not give rise to stray flux and thus would lead to
smaller energy consumption in devices.  Also half-metallic
ferrimagnetism (HMFi) is highly desirable in the absence of HMA
since such compounds would yield lower total spin moments than the
corresponding ferromagnets. Well-known HMFi are the perfect
Heusler compounds FeMnSb and Mn$_2$VAl \cite{deGroot2}. We will
present in this section another route to half-metallic
ferrimagnetism based on antisites created by the migration of
Cr(Mn) atoms at Co sites in the
 case of Co$_2$CrAl, Co$_2$CrSi, Co$_2$MnAl and Co$_2$MnSi alloys.

\begin{figure}
\includegraphics[width=\textwidth]{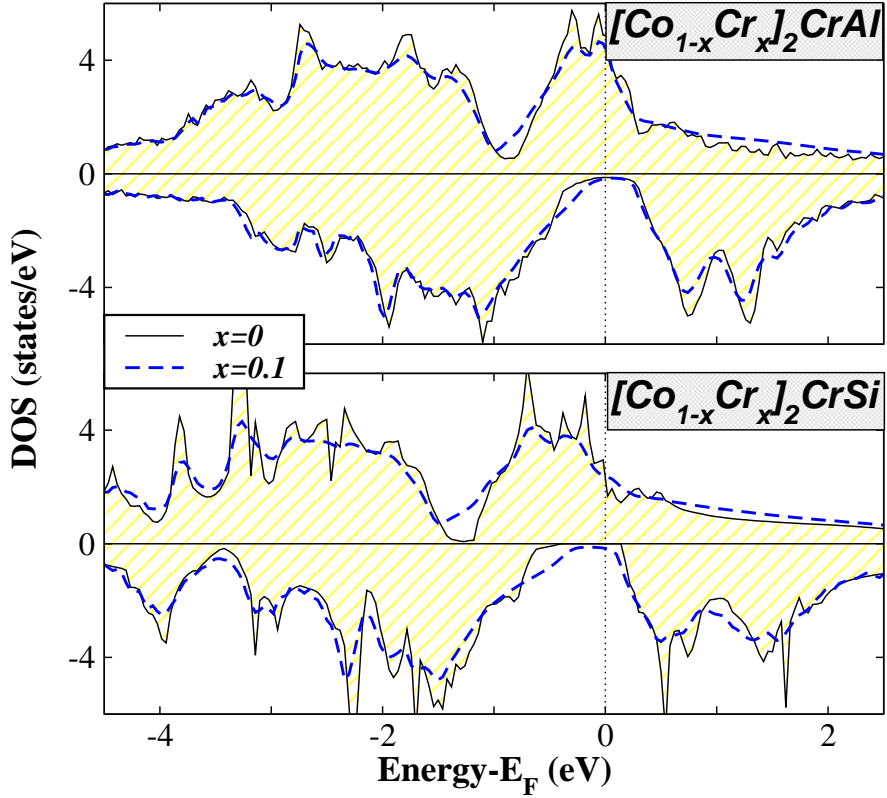}
\caption{(Color online) Total density of states (DOS) as a
function of the concentration $x$  for the
[Co$_{1-x}$Cr$_{x}$]$_2$CrAl (upper panel) and
[Co$_{1-x}$Cr$_{x}$]$_2$CrSi (lower panel) compounds.
\label{fig_HMFi}}
\end{figure}

\begin{table}
\caption{Atom-resolved spin magnetic moments for the
[Co$_{1-x}$Cr$_x$]$_2$CrAl, [Co$_{1-x}$Cr$_x$]$_2$CrSi,
[Co$_{1-x}$Mn$_x$]$_2$MnAl and [Co$_{1-x}$Mn$_x$]$_2$MnSi
compounds (moments have been scaled to one atom). The two last
columns are the total spin moment (Total) in the unit cell
calculated as $2\times [(1-x)*m^{Co}+x*m^{Cr \:\mathrm{or}\:
Mn(imp)}]+m^{Cr(Mn)}+m^{Al\:\mathrm{or}\:Si}$ and the ideal total
spin moment predicted by the Slater-Pauling rule for half-metals
(see reference \cite{GalaFull}). With Cr(imp) or Mn(imp) we denote
the Cr(Mn) atoms sitting at perfect Co sites. \label{table_HMFi} }
\begin{tabular}{lcccccc}\hline\noalign{\smallskip}
 \multicolumn{7}{c}{[Co$_{1-x}$Cr$_x$]$_2$CrAl} \\
 $x$ & Co   & Cr(imp) & Cr & Al & Total &Ideal \\
  0    &  0.73 & -- &  1.63 & -0.09&3.00  & 3.00 \\

0.05 & 0.71 & -1.82 & 1.62 & -0.09 & 2.70 & 2.70 \\

0.1 & 0.69 & -1.85 & 1.61 & -0.08 & 2.40 &  2.40 \\

0.2 & 0.64 & -1.87 & 1.60 & 0.06 & 1.80 & 1.80 \\ \hline

 \multicolumn{7}{c}{[Co$_{1-x}$Cr$_x$]$_2$CrSi} \\
 $x$ & Co   & Cr(imp) & Cr & Si & Total &Ideal \\

0 & 0.95 & -- & 2.17 & -0.06 & 4.00 &  4.00 \\

0.05 & 0.93 & -1.26 & 2.12 & -0.06 & 3.70 & 3.70 \\

0.1 & 0.91 & -1.26 & 2.07 & -0.05 & 3.40 & 3.40 \\

0.2 & 0.87 & -1.25 & 1.96 & -0.04 & 2.80 & 2.80 \\
\hline
\multicolumn{7}{c}{[Co$_{1-x}$Mn$_x$]$_2$MnAl} \\
 $x$ & Co   & Mn(imp) & Mn & Al & Total & Ideal \\
  0    &  0.68  & -- & 2.82 & -0.14 & 4.04 & 4.00 \\

0.05 & 0.73 & -2.59 & 2.82 & -0.13 & 3.81 & 3.80 \\

0.1 & 0.78 & -2.49 & 2.83 & -0.12 & 3.61 & 3.60 \\

0.2 & 0.84 & -2.23 & 2.85 & -0.09 & 3.20 &3.20 \\

 \hline
 \multicolumn{7}{c}{[Co$_{1-x}$Mn$_x$]$_2$MnSi} \\

$x$ & Co   & Mn(imp) & Mn & Si & Total &Ideal \\

0.0&  0.98  & -- & 3.13 & -0.09 & 5.00 & 5.00\\

0.05& 0.99 & -0.95 & 3.09 & -0.08 & 4.80 & 4.80\\

0.1 & 0.99 & -0.84 & 3.06 & -0.07 & 4.60& 4.60 \\

0.2& 0.97 & -0.70 & 2.99 & -0.05 & 4.20 & 4.20 \\
\noalign{\smallskip}\hline

\end{tabular}
\end{table}

We will start our discussion from  the Cr-based alloys and using
Co$_2$CrAl and Co$_2$CrSi as parent compounds we create a surplus
of Cr atoms which sit at the perfect Co sites. In figure
\ref{fig_HMFi}  we present the total density of states (DOS) for
the [Co$_{1-x}$Cr$_{x}$]$_2$CrAl  and [Co$_{1-x}$Cr$_{x}$]$_2$CrSi
alloys for concentrations $x$= 0 and 0.1, and in table
\ref{table_HMFi} we have gathered the spin moments for the two
compounds under study. We will start our discussion from the DOS.
The perfect compounds show a gap in the minority-spin band and the
Fermi level falls within this gap and thus the compounds are
half-metals. When the sp atom is Si instead of Al the gap is
larger due to the extra electron which occupies majority states of
the transition metal atoms \cite{GalaFull} and increases the
exchange splitting between the majority occupied and the minority
unoccupied states. This electron increases the Cr spin moment by
$\sim$0.5 $\mu_B$ and the moment of each Co atom by $\sim$0.25
$\mu_B$ about. The Cr and Co majority states form a common band
and the weight at the Fermi level is mainly of Cr character. The
minority occupied states are mainly of Co character. When we
substitute Cr for Co, the effect on the atomic DOS of the Co and
Cr atoms at the perfect sites is marginal. The DOS of the impurity
Cr atoms has a completely different form from the Cr atoms at the
perfect sites due to the different symmetry of the site where they
sit. But although Cr impurity atoms at the antisites induce
minority states within the gap, there is still a tiny gap and the
Fermi level falls within this gap keeping the half-metallic
character of the parent compounds even when we substitute 20\% of
the Co atoms by Cr ones.

The discussion above on the conservation of the half-metallicity
is confirmed when we compare the calculated total moments in table
\ref{table_HMFi}  with the values predicted by the Slater Pauling
rule for the ideal half-metals (the total spin moment in $\mu_B$
is the number of valence electrons in the unit cell minus 24)
\cite{GalaFull}. Since Cr is lighter than Co, substitution of Cr
for Co decreases the total number of valence electrons and the
total spin moment should also decrease. The interesting point is
the way that the reduction of the total spin moment is achieved.
Co and Cr atoms at the perfect sites show a small variation of
their spin magnetic moments with the creation of defects and the
total spin moment is reduced due to  the antiferromagnetic
coupling between the Cr impurity atoms and the Co and Cr ones at
the ideal sites, which would have an important negative
contribution to the total moment as confirmed by the results in
table \ref{table_HMFi}. Thus the Cr-doped alloys are half-metallic
ferrimagnets and their total spin moment is considerable smaller
than the perfect half-metallic ferromagnetic parent compounds; in
the case of [Co$_{0.8}$Cr$_{0.2}$]$_2$CrAl it decreases down to
1.8 $\mu_B$ from the 3 $\mu_B$ of the perfect Co$_2$CrAl alloy.
Here we have to mention that if also Co atoms migrate to Cr sites
(case of atomic swaps) the half-metallity is lost, as it was shown
by Miura et al. \cite{Miura}, due to the energy position of the Co
states which have migrated at Cr sites.

\begin{figure}
\begin{center}
\includegraphics[width=\textwidth]{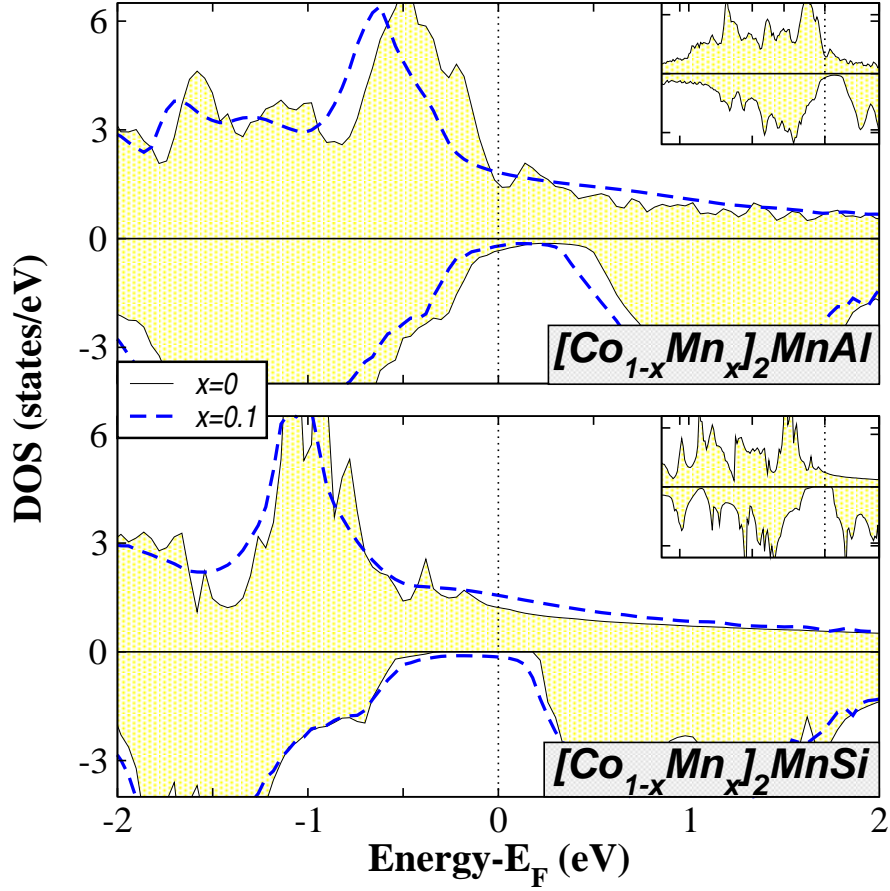}
\end{center}
\caption{(Color online) Total density of states (DOS) around the
gap region for the [Co$_{1-x}$Mn$_{x}$]$_2$MnAl
 and  [Co$_{1-x}$Mn$_{x}$]$_2$MnSi alloys as a function of the
concentration $x$ : we denote $x=0$ with the solid grey line with
shaded region and $x=0.1$ with a dashed thick blue line. In the
insets we present the DOS for a wider energy region and for a
$x=0$ concentration.} \label{fig_HMFi2}
\end{figure}

In the rest of this section we will present results on the
[Co$_{1-x}$Mn$_x$]$_2$MnZ compounds varying the sp atom, Z, which
is one of Al or Si.  We have taken into account five different
values for the concentration $x$; $x$= 0, 0.025, 0.05, 0.1, 0.2.
In figure \ref{fig_HMFi2} we have drawn the total density of
states (DOS) for both families of compounds under study and for
two different values of the concentration $x$ : the perfect
compounds ($x$=0) and for one case with defects, $x$= 0.1. In the
case of the perfect Co$_2$MnSi compound there is a real gap in the
minority spin band and the Fermi level falls within this gap and
this compound is a perfect half-metal. Co$_2$MnAl presents in
reality a region of tiny minority-spin DOS instead of a real gap,
but the spin-polarization at the Fermi level only marginally
deviates from the ideal 100\% and this alloy
 can be also considered as half-metal. These results agree with
previous electronic structure calculations on these compounds
\cite{GalaFull,Picozzi,JAP,APL,Richter}. When we create a surplus
of Mn atoms which migrate at sites occupied by Co atoms in the
perfect alloys, the gap persists and both compounds retain their
half-metallic character as occurrs also for the Cr-based alloys
presented above.   Especially for Co$_2$MnSi, the creation of Mn
antisites does not alter the width of the gap and the
half-metallicity is extremely robust in these alloys with respect
to the creation of Mn antisites. The atomic spin moments show
behavior similar to the Cr alloys as can be seen in table
\ref{table_HMFi} and the spin moments of the Mn impurity atoms are
antiferromagnetically coupled to the spin moments of the Co and Mn
atoms at the perfect sites resulting to the desired half-metallic
ferrimagnetism.

\section{A possible route to half-metallic antiferromagnetism \label{sec4}}

Now, we will present a way to use the defects in perfect
half-metallic ferrimagnets in order to create a half-metallic
antiferromagnetic material:  the doping with Co of the Mn$_2$VAl
and Mn$_2$VSi alloys which are well known to be HMFi. The
importance of this route stems from the existence of Mn$_2$VAl in
the Heusler $L2_1$ phase as shown by several groups
\cite{itoh1,itoh2,itoh3}. Each Mn atom has a spin moment of around
-1.5 $\mu_B$ and V atom a moment of about 0.9 $\mu_B$
\cite{itoh1,itoh2,itoh3}.

\begin{figure}
\begin{center}
\includegraphics[width=\textwidth]{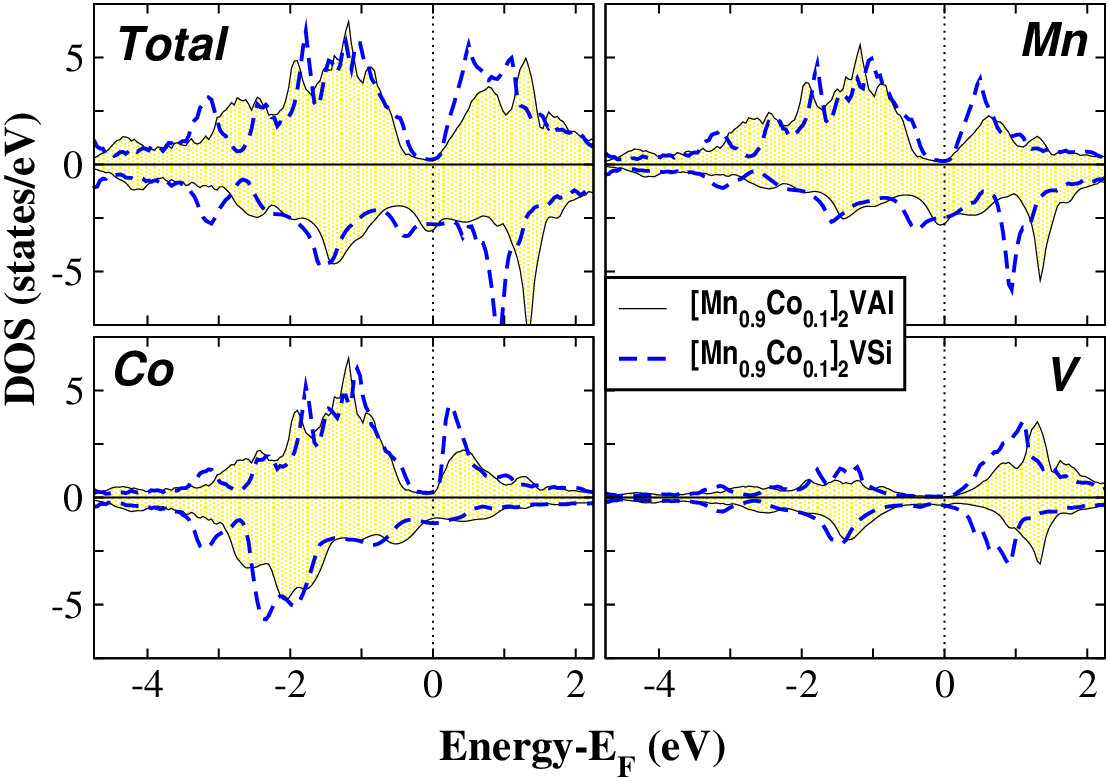}
\end{center}
\caption{(Color online) Total and atom-resolved DOS for the
[Mn$_{0.9}$Co$_{0.1}$]$_2$VAl and [Mn$_{0.9}$Co$_{0.1}$]$_2$VSi
compounds. Note that the atomic DOS's have been scaled to one
atom. Positive values of DOS correspond to the spin-up (minority)
electrons while negative values correspond to the spin-down
(majority) electrons. \label{fig_HMA} }
\end{figure}

 All theoretical studies on
Mn$_2$VAl agree on the half-metallic character with a gap at the
spin-up band instead of the spin-down band as for the other
half-metallic Heusler alloys \cite{GalaFull,Mn2VZ-2,Weht1,Weht2}.
Prior to the presentation of our results we have to note that due
to the Slater-Pauling rule \cite{GalaFull}, these compounds with
less than 24 valence electrons have negative total spin moments
and the gap is located at the spin-up band. Moreover the spin-up
electrons correspond to the minority-spin electrons and the
spin-down electrons to the majority electrons contrary to the
other Heusler alloys \cite{GalaFull}. We have substituted Co for
Mn in Mn$_2$V(Al or Si) in a random way and in figure
\ref{fig_HMA} we present the total and atom-resolved density of
states (DOS) in [Mn$_{1-x}$Co$_x$]$_2$VAl (solid black line) and
[Mn$_{1-x}$Co$_x$]$_2$VSi (blue dashed line) alloys for $x$=0.1.
The perfect compounds show a region of low spin-up DOS (we will
call it a ``pseudogap'') instead of a real gap. Upon doping the
pseudogap at the spin-up band persists and the quaternary alloys
keep the half-metallic character of the perfect Mn$_2$VAl and
Mn$_2$VSi compounds.  Co atoms are strongly polarized by the Mn
atoms since they occupy the same sublattice and they form Co-Mn
hybrids which afterwards interact with the V and Al or Si states
\cite{GalaFull}. The spin-up Co states form a common band with the
Mn ones and the spin-up DOS for both atoms has similar shape. Mn
atoms have less weight in the spin-down band since they
accommodate less charge than the heavier Co atoms.

In table \ref{table_HMA} we have gathered the total and
atom-resolved spin moments for all the Co-doped compounds as a
function of the concentration. We have gone up to a concentration
which corresponds to 24 valence electrons in the unit cell, thus
up to $x$=0.5 for the [Mn$_{1-x}$Co$_x$]$_2$VAl and x=0.25 for the
[Mn$_{1-x}$Co$_x$]$_2$VSi alloys. In the last column we have
included the total spin moment predicted by the Slater-Pauling
rule for the perfect half-metals \cite{GalaFull}. A comparison
between the calculated and ideal total spin moments reveals that
all the compounds under study are half-metals with very small
deviations due to the existence of a pseudogap instead of a real
gap. Exactly for 24 valence electrons the total spin moment
vanishes as we will discuss in the next paragraph. Co atoms have a
spin moment parallel to the V one and antiparallel to the Mn
moment, and thus the compounds retain their ferrimagnetic
character. As we increase the concentration of the Co atoms in the
alloys, each Co  has more Co atoms as neighbors, it hybridizes
stronger with them and its spin moment increases while the spin
moment of the Mn atom decreases (these changes are not too
drastic). The sp atoms have a spin moment antiparallel to the Mn
atoms as already discussed in reference \cite{Mn2VZ}.

\begin{table}
\caption{Atom-resolved spin magnetic moments for the
[Mn$_{1-x}$Co$_x$]$_2$VAl and [Mn$_{1-x}$Co$_x$]$_2$VSi compounds
(moments have been scaled to one atom). The two last columns are
the total spin moment (Total) in the unit cell calculated as
$2\times [(1-x)*m^{Mn}+x*m^{Co}]+m^V+m^{Al\:or\:Si}$ and the ideal
total spin moment predicted by the Slater-Pauling rule for
half-metals (see reference \cite{GalaFull}). The lattice constants
have been chosen 0.605 nm for Mn$_2$VAl and 0.6175 for Mn$_2$VSi
for which both systems are half-metals (see reference
\cite{Mn2VZ}) and have been kept constant upon Co doping. }
\label{table_HMA}
 \begin{tabular}{lcccccc}
 \hline\noalign{\smallskip}
 \multicolumn{7}{c}{[Mn$_{1-x}$Co$_x$]$_2$VAl} \\
 $x$ & Mn   & Co & V & Al & Total &Ideal \\
  0    & -1.573  & -- & 1.082  & 0.064 & -2.000 & -2.0 \\

0.05 & -1.580 & 0.403 & 1.090 & 0.073 & -1.799 & -1.8 \\

0.1 & -1.564 & 0.398 & 1.067 & 0.069 & -1.600 & -1.6 \\

0.3 & -1.484 & 0.456 & 0.953 & 0.047 & -0.804 & -0.8 \\

0.5& -1.388 & 0.586 & 0.782 & 0.019 & $\sim$0 & 0\\ \hline

\multicolumn{7}{c}{[Mn$_{1-x}$Co$_x$]$_2$VSi} \\
 $x$ & Mn   & Co & V & Si & Total &Ideal \\

0 &  -0.960 & -- & 0.856 & 0.063&-1.000 &  -1.0 \\

0.05 & -0.944 & 0.749 & 0.860 & 0.059 & -0.800 & -0.8 \\

0.1 & -0.925 & 0.819 & 0.847 & 0.054 & -0.600 & -0.6 \\

0.2 & -0.905& 0.907 & 0.839 & 0.046 & -0.201 & -0.2 \\

0.25&-0.899 & 0.935 & 0.839 & 0.041 & $\sim$0 & 0 \\
\noalign{\smallskip}\hline
\end{tabular}

\end{table}

The most interesting point in this substitution procedure is
revealed when  we increase the Co concentration to a value
corresponding to 24 valence electrons in the unit cell, thus  the
[Mn$_{0.5}$Co$_{0.5}$]$_2$VAl and [Mn$_{0.75}$Co$_{0.25}$]$_2$VSi
alloys. The Slater-Pauling rule predicts for these compounds a
zero total spin moment in the unit cell and  the electrons
population is equally divided between the two spin-bands. Our
first-principles calculations reveal that this is actually the
case. The interest arises from the fact that although the total
moment is zero, these two compounds are made up from  strongly
magnetic components. Mn atoms have a mean spin moment of
$\sim$-1.4 $\mu_B$ in [Mn$_{0.5}$Co$_{0.5}$]$_2$VAl and $\sim$-0.9
$\mu_B$ in [Mn$_{0.75}$Co$_{0.25}$]$_2$VSi. Co and V have spin
moments antiferromagnetically coupled to the Mn ones which for
[Mn$_{0.5}$Co$_{0.5}$]$_2$VAl are  $\sim$0.6 and $\sim$0.8
$\mu_B$, respectively, and for [Mn$_{0.75}$Co$_{0.25}$]$_2$VSi
$\sim$0.9 and $\sim$0.8 $\mu_B$. Thus these two compounds are
half-metallic fully-compensated ferrimagnets or as they are best
known in literature half-metallic antiferromagnets.

\section{Vacancies \label{sec5}}

\begin{figure}
\includegraphics[width=\textwidth]{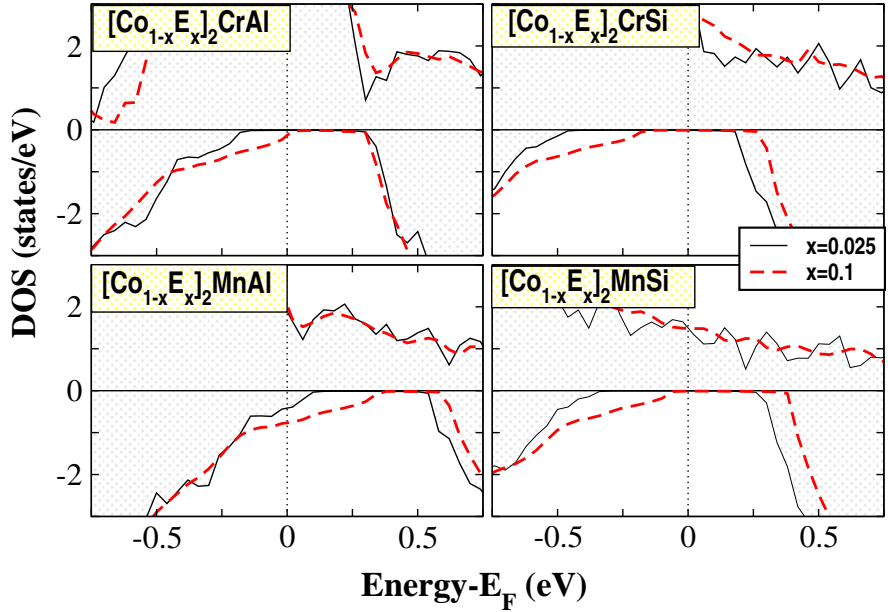}
\caption{(Color online) Total density of states (DOS) as a
function of the concentration $x$  for the four studied alloys
[Co$_{1-x}$E$_{x}$]$_2$YZ where Y stands for Cr and Mn, Z for Al
and Si and  E the vacant site. \label{vac_fig}}
\end{figure}

Finally we will also shortly discuss a special case: the occurence
of vacancies \cite{PSS-RRL2}. In experiments a vacancy can appear
in all possible sites and we will start our discussion by the case
of vacancies (E) appearing at the sites occupied by Co atoms at
the perfect compounds. In figure \ref{vac_fig} we have gathered
the total density of states (DOS) for the four studied compounds,
[Co$_{1-x}$E$_x$]$_2$YZ (Y is Cr or Mn and Z is Al or Si) when we
replace 2.5\%\ (x=0.025) and 10\%\ (x=0.10) of the Co atoms.
Although there is no visible effect on the properties of the gap
with respect to the perfect compound \cite{APL,SSC} when $x$ is as
small as 0.025,  as we increase $x$ to 0.1 impurity states located
at the left edge of the gap appear.  The vacancy carries very
small charge with a vanishing DOS and the  induced minority states
are mainly located at the Co sites (Mn and Si atoms have a very
small DOS at the same energy range as a result of the polarization
of these Co-located vacancy-induced minority-states). These states
result in a considerable shrinking of the width of the gap of the
order of 0.25-0.3 eV.  In the case where the Fermi level is close
to the left edge of the gap as in the Co$_2$MnAl alloy the
shrinking of the gap results in the loss of half-metallicity (the
spin-polarization is reduced to $\sim$ 40\% for $x=0.1$).

In the case when the  vacancies occur at the Y site occupied in
the perfect alloys by Cr or Mn atoms and the Z site occupied by Al
or Si, they again induce minority states within the left edge of
the gap leading to a shrinking of its width as in the case
discussed in the previous paragraph. But the effect is much more
mild leading to a shrinking of the gap by 0.1 eV approximately
when they substitute Y-type atoms and even less when they
substitute the sp atoms. Moreover, not only the reduction of the
width of the gap is smaller, but also the intensity of the
minority-spin induced DOS within the gap is smaller with respect
to the case discussed in the previous paragraph. This difference
in the behavior of the vacancy-induced minority states has its
routes to the fact that the gap is created between states
exclusively localized at the Co sites which due to symmetry
reasons do not couple to Cr(Mn) or Al(Si) orbitals. When the
vacancy substitutes Co atoms the effect is much more intense
around the Fermi level than when it substitutes Cr(Mn) or Al(Si)
atoms which have almost zero weight around the gap. This is also
reflected on the change of the behavior of the atomic spin moments
(see reference \cite{PSS-RRL2}) for an extended discussion of spin
moments).

\section{Summary and outlook \label{sec6}}

In this chapter we have reviewed our results on the defects in
half-metallic Heusler alloys. Firstly we have studied the effect
of doping and disorder on the magnetic properties of the
Co$_2$MnAl(Si)  full-Heusler alloys. Doping simulated by the
substitution of Cr and Fe for Mn overall keeps the
half-metallicity.  Both disorder and doping have little effect on
the half-metallic properties of Co$_2$MnSi   and it keeps a high
degree of spin-polarization. Co$_2$MnAl presents a region of low
minority density of states instead of a real gap. Doping  keeps
the half-metallicity of Co$_2$MnAl while  disorder simulated by
excess of either the Mn or Al atoms completely destroys the almost
perfect spin-polarization of the perfect compound contrary to
Co$_2$MnSi.

Afterwards, we have studied the effect of defects-driven
appearance of half-metallic ferrimagnetism in the case of the
Co$_2$Cr(Mn)Al(Si) Heusler alloys.  More precisely, based on
first-principles calculations we have shown that when we create
Cr(Mn) antisites at the Co sites, these impurity Cr(Mn) atoms
couple antiferromagnetically with the Co and the Cr(Mn) atoms at
the perfect sites while keeping the half-metallic character of the
parent compounds. This is a promising alternative way to create
robust half-metallic ferrimagnets, which are crucial for
magnetoelectronic applications.

\noindent Moreover we have also studied the effect of doping the
half-metallic ferrimagnets Mn$_2$VAl and Mn$_2$VSi. Co
substitution for Mn keeps the half-metallic character of the
parent compounds and when the total number of valence electrons
reaches the 24, the total spin moment vanishes as predicted by the
Slater-Pauling rule and  the ideal half-metallic
antiferromagnetism is achieved. Thus we can create half-metallic
antiferromagnets simply by introducing Co atoms in the Mn$_2$VAl
and Mn$_2$VSi half-metallic ferrimagnets. Since crystals and films
of both Mn$_2$VAl and Co$_2$VAl alloys have been grown
experimentally, such a compound maybe is feasible experimentally.

Finally, we have studied the effect of vacancies in half-metallic
Heusler alloys. We have shown using ab-initio electronic structure
calculations that the occurrence of vacancies at the sites
occupied by Co atoms in the perfect compounds seriously affects
the stability of their half-metallic character and the
minority-spin gap is rapidly shrinking. On the other hand
vacancies at the other sites do not have an important impact on
the electronic properties of the half-metallic full-Heusler
compounds. Thus it is crucial for spintronic applications to
prevent creation of vacancies during the growth of the
half-metallic full-Heusler alloys used as electrodes in the
magnetoelectronic devices.

Although we have presented several aspects of defects in Heusler
alloys, many more calculations and experiments are needed; the aim
is to find systems, which either do not lead to states in the gap
(like the defects-driven half-metallic ferrimagnets presented in
sections \ref{sec3} and \ref{sec4}) or systems with particularly
high defect formation energies or sufficiently low annealing
temperatures (we have not discussed the energetics of defects in
this contribution). Equally important for realistic applications
is the control of surface and interface states in the gap, the
latter are in particular important for interfaces to
semiconductors, which we have not been examined in this chapter.


\end{document}